\documentclass{aa}
\usepackage{graphicx}
\usepackage{txfonts}

\usepackage{natbib} 
\bibpunct{(}{)}{;}{a}{}{,}

\begin{document}

\title{An uncombed inversion of multi-wavelength observations reproducing the Net Circular Polarization in a sunspots' penumbra}
\author{C. Beck\inst{1,2}}
\titlerunning{An uncombed inversion of multi-wavelength observations}
\authorrunning{C. Beck}  
\offprints{C. Beck}
\institute{Instituto de Astrof\'isica de Canarias
  (CSIC)
  \and Departamento de Astrof\'isica, Universidad de La Laguna
}
\date{Received xxx; accepted xxx}
\abstract{The penumbra of sunspots shows a complex magnetic field topology
  whose three-dimensional organization is still under debate after more than a
  century of investigation.}{I derived a geometrical model of the penumbral
  magnetic field topology from an uncombed inversion setup that aimed at
  reproducing the Net Circular Polarization (NCP) of simultaneous spectra in
  near-infrared (IR; 1.56\,$\mu$m) and visible (VIS; 630\,nm)
  spectral lines.}{I inverted the co-spatial spectra of five photospheric
  lines with a model that mimicked vertically interlaced magnetic fields with
  two distinct components, labeled background field and flow channels because
  of their characteristic properties (flow velocity, field inclination). The
  flow channels were modeled as a perturbation of the constant background
  field with a Gaussian shape using the SIRGAUS code (Bellot Rubio 2003). The location and extension of the Gaussian perturbation in the optical depth scale retrieved by the inversion code was then converted to a geometrical height scale. With the estimate on the geometrical size of the flow channels, I investigated the relative amount of magnetic flux in the flow channels and the background field atmosphere.}{The uncombed model is able to reproduce the NCP well on the limb side of the spot and worse on the center side; the VIS lines are better reproduced than the near-IR lines. I find that the Evershed flow happens along nearly horizontal field lines close to the solar surface given by optical depth unity. The magnetic flux that is related to the flow channels makes up about 20-50\% of the total magnetic flux in the penumbra.}{The gradients obtainable by a Gaussian perturbation are too small for a perfect reproduction of the NCP in the IR lines with their small formation height range, where a step function seems to be required. Two peculiarities of the observed NCP, a sign change of the NCP of the VIS lines on the center side and a ring structure around the umbra with opposite signs of the NCP in the \ion{Ti}{i} line at 630.37\,nm and the \ion{Fe}{i} line at 1565.2\,nm deserve closer attention in future modeling attempts. The large fraction of magnetic flux related to the flow channel component could allow to replenish the penumbral radiative losses in the flux tube picture.}
\maketitle
\section{Introduction}
It has not been possible up to now to unambiguously determine the
topology of the magnetic fields inside the penumbra of sunspots directly from
spectroscopic or spectropolarimetric observations due to the many different possibilities of interpreting the data. The observations have
provided some boundary conditions, like for instance the presence of the
Evershed flow \citep{evershed1909}, an almost radial orientation of the
magnetic field lines, a radial decrease of magnetic field strength, and a
radial increase of the field inclination to the local vertical
\citep[e.g.,][]{lites+etal1993,westendorp+etal2001,borrero+bellot2002,solanki2003,bellot+etal2004,langhans+etal2005,beck2008,schlichemaier2009}.
With increasing spatial resolution, the Evershed flow could be definitely be
shown to be related to the penumbral filaments
\citep[][]{shine+etal1994,tritschler+etal2004,langhans+etal2005,rimmele+marino2006}.
The penumbral filaments have a very peculiar appearance in data of highest
spatial resolution, a dark core flanked by two lateral brightenings
\citep{scharmer+etal2002,suetterlin+etal2004}, with indications that the flow
velocity is highest in the darkest part
\citep{bellot+etal2005,langhans+etal2007,bellotrubio+etal2007}. The small
scales of this internal structure of penumbral filaments presumably
contributed to the fact that conflicting results on the relation between flow
velocity and intensity were found in several previous studies
\citep[e.g.,][]{degenhardt+wiehr1991,title+etal1993,lites+etal1993,shine+etal1994,schliche+luis+ali2005,langhans+etal2005}.
It has not been possible to put together all the information on the penumbra into an unique commonly accepted model, also thanks to another source of information which spectropolarimetric observations provide, the Net Circular Polarization (NCP), a measure of the asymmetry of the Stokes $V$ polarization signal. The penumbra of sunspots shows one of the largest NCP values of all solar structures \citep{illing+etal1974,auer+heasley1978,henson+kemp1984,almeida+lites1992,mueller+etal2006,ichimoto+etal2008}. Since the NCP is related to gradients of magnetic fields strength and velocity along the line of sight (LOS) \citep[e.g.,][]{skumanich+lites1987a,almeida+lites1992}, it is a crucial source of information on the three-dimensional organization of the penumbral magnetic fields. Unfortunately, gradients can come in even more different shapes than constant values, opening up even more possible configurations to reproduce the observations. 

One of the suggested configurations for the three-dimensional organization of the penumbral magnetic fields is the so-called ``uncombed'' penumbra \citep{solanki+montavon1993}, where less inclined field lines wind around horizontal flow channels \citep[see for instance][]{borrero+etal2008}. This configuration has been taken up for modeling \citep[][]{thomas+montesinos1993,schliche+jahn+schmidt1998,schlichenmaier+collados2002,mueller+etal2002,borrero+etal2004,mueller+etal2006}\footnote{Thomas \& Montesinos actually predates the paper of Solanki \& Montavon.} and inversion schemes since it is one of the few approaches that reproduces the observed NCP \citep[e.g.,][]{pillet2000,borrero+etal2006,beckthesis2006,jurcak+etal2007,borrero+etal2007}. Another approach reproducing the observed NCP is the so-called micro-structured magnetic atmosphere hypothesis \citep[MISMA;][]{almeida1998,almeida2001,almeida2005} that proposes a fine structure of the magnetic field at scales of a few km. On an average over the scales that are typical for the spatial resolution of even the most recent observations, the MISMA can be represented by two components similar to the uncombed model \citep{almeida2005}, but with additional properties that come from its substructure. 

Whereas the uncombed models were derived in close connection to, or in some cases, from spectropolarimetric observations of sunspots, some other explanations for the penumbral structure have been brought forward from a more theoretical point of view. \citet{thomas+weiss2004} suggested the effect of magnetic flux pumping as acting agent of the penumbral fine-structure. This provides a driver for the organization of the penumbral magnetic fields, but gives, however, no description of the organization itself. \citet{scharmer+spruit2006} and \citet{spruit+scharmer2006} proposed the existence of field-free gaps with convective motions below the visible surface to balance the radiative energy losses of the penumbra, but it has not been shown that this model finally reproduces the observed NCP. The numerical 3-D MHD simulations described by \citet{rempel+etal2009,rempel+etal2009a} showed the effects of magneto-convection in the simulated sunspot, i.e., convection cells oriented and shaped by the direction of the magnetic field lines rather than completely field-free gaps, but again the cross-check of the resulting spectra vs.~spectropolarimetric observations is still missing besides an initial attempt by \citet{borrero+etal2010}. 

In this contribution, I investigate the NCP in simultaneous spectra in the
visible (VIS; 630\,nm) and near-infrared (IR; 1.56\,$\mu$m) using
an uncombed inversion setup that includes gradients along the LOS in field strength and velocity. The observations are described in Sect.~\ref{sect_obs}. The inversion method is explained in Sect.~\ref{sect_analysis}, its results are described in Sect.~\ref{sect_results}. Section \ref{sect_disc} discusses the findings, while Sect.~\ref{sect_concl} gives the conclusions. Appendix \ref{inicomp} shows the differences between initial and best-fit model atmospheres, Appendix \ref{prof_examples} several examples of observed spectra with the corresponding best-fit profiles.
\section{Observations, data reduction and data alignment\label{sect_obs}}
The observations were taken on 2003 Aug 9 with the POlarimetric LIttrow Spectrograph \citep[POLIS;][]{beck+etal2005b} and the Tenerife Infrared Polarimeter \citep[TIP;][]{martinez+etal1999} at the German Vacuum Tower Telescope (VTT) in Iza{\~n}a, Tenerife, Spain. These instruments observed the Stokes vector at 1565\,nm and 630\,nm, respectively. Both instruments were fed simultaneously using an achromatic 50-50 beamsplitter. The setup and the data are described in detail in \citet{beck+etal2006d} and \citet{beck2008}; here I only used one of the two observations described in the latter. The data were reduced with the respective calibration routines for both instruments, including the correction for the instrumental polarization of the VTT \citep[e.g][]{beck+etal2005a,beck+etal2005b}. The spectra were aligned and brought to an identical spatial sampling as described in the Appendix of \citet{beck+etal2006d}. The spatial resolution was estimated to be around 1$^{\prime\prime}$ \citep{beck+etal2006d}. In the inversion of the spectra, the transitions of the spectral lines and the adopted rest wavelengths were identical to those given in Table 1 of \citet{beck2008}, but without the two weakest lines (\ion{Fe}{i} at 1564.74\,nm and \ion{Fe}{i} at 630.35\,nm).   
\begin{figure}
\resizebox{8.8cm}{!}{\hspace*{.5cm}
\begin{minipage}{7.7cm}
\resizebox{7.7cm}{!}{\includegraphics{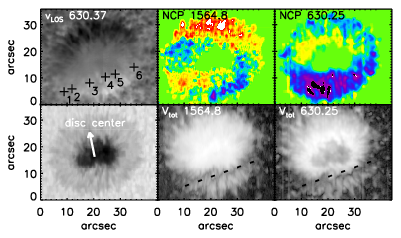}}$ $\\
\end{minipage}
\begin{minipage}{1.cm}
\resizebox{.95cm}{!}{\includegraphics{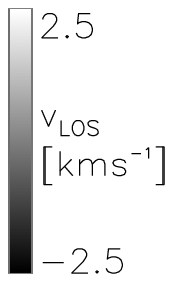}}\vspace*{.03cm}\\
\resizebox{.95cm}{!}{\includegraphics{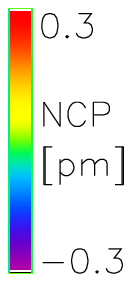}}$ $\\
\end{minipage}}\vspace*{.2cm}
\caption{Overview of the observation. {\em Bottom row, left to right}:
  continuum intensity in IR, integrated absolute Stokes $V$ signal of 1564.8\,nm, same for 630.25\,nm. {\em Top row, left to right}: line-core velocity of
  630.37\,nm, NCP of 1564.8\,nm, same for 630.25\,nm. Tick marks are in
  arcsec. The {\em black dashed} line denotes the location of a cut along the
  neutral line of Stokes $V$. The {\em white arrow} points towards disc
  center. The {\em black crosses} and the corresponding numbers in the
    map of  630.37\,nm denote the locations of the profiles shown in Figs.~\ref{prof1} to \ref{prof3}. \label{fig1}}
\end{figure}

The sunspot NOAA 10425 was located at an heliocentric angle of around 30$^\circ$. Figure \ref{fig1} shows the IR continuum intensity map, the integrated absolute Stokes $V$ signal of the two more magnetic sensitive IR and VIS lines (1564.8\,nm, 630.25\,nm), the line-core velocity of the weak \ion{Ti}{i} line at 630.37\,nm, and the NCP of the lines at 630.25\,nm and 1564.8\,nm. The Evershed effect and its filamentary structure can be seen in the velocity map of the \ion{Ti}{i} line. This line is only slightly sensitive to magnetic fields, and hence allows to recover reliable line-core velocities also in the umbra where the other lines split. The two NCP maps show the symmetry pattern described by \citet[][2006]{mueller+etal2002}: the VIS is symmetric to the line of symmetry through sunspot center and disc center, the IR anti-symmetric with twice the frequency. Comparison of the NCP with the integrated $V$ signal shows that the NCP is not changing its properties in the neutral line of Stokes $V$ on the limb side even if the absolute $V$ signal reduces strongly. 

\section{Data analysis\label{sect_analysis}}
The co-spatial spectra of all observed VIS and IR lines were first inverted
simultaneously with the standard version of the SIR code
\citep{cobo+toroiniesta1992}. This analysis used two magnetic field
  components in each pixel inside the penumbra where the field properties were
  assumed to be constant with optical depth (termed  ``{\it 2C inversion}'' in the following). The inversion setup and its results are described in detail in \citet{beck2008}. This inversion setup is unable to produce any NCP in its resulting synthetic spectra, but is, however, important both for comparison and as initial model for the more complex ``{\it Gaussian inversion}'' described below. The 2C inversion is not reflecting the topology of the magnetic fields in the penumbra satisfyingly, because it corresponds to two magnetic field components with different field inclinations that are horizontally separated. To explain the observed NCP, a vertical interlacing of the fields has to be assumed. In the ``uncombed'' picture, the geometry is commonly modeled as horizontal flow channels of a limited width around which the less inclined field lines wind. A LOS that passes through such a configuration will thus encounter first the less inclined field, then at some depth the flow channel, and eventually again the background (bg) field, depending on the opacity of the flow channel (fc). One way to mimic such a geometry in an inversion of spectra is to introduce a localized perturbation at some optical depth in the atmospheric stratification.
\begin{figure}
\centerline{\resizebox{8.8cm}{!}{\includegraphics{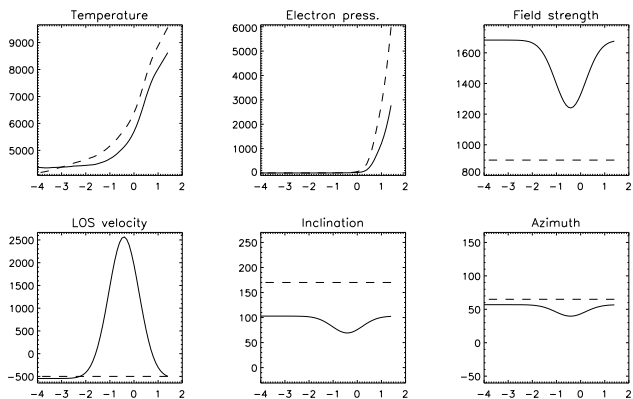}}}
\caption{Example of the model atmosphere for the Gaussian inversion. {\em Top
    row, left to right}: temperature in Kelvin, electron pressure in 
    dyn\,cm$^{-2}$, field strength in Gauss. {\em Bottom row}: line-of-sight
  velocity in m\,s$^{-1}$, field inclination to the LOS in degree, field
  azimuth in degree. {\em Dashed}: initial values of the background field atmosphere,
  {\em solid}: final best-fit background field with perturbation added. The
    values are given vs.~logarithmic optical depth, the final background field values can be simply read off from the values at $\log\tau =-4$ since they are constant with depth.\label{gausmodel}}
\end{figure}

To this extent, the SIRGAUS code \citep{bellot2003} was used. SIRGAUS is
  based on the SIR code and allows to add a Gaussian perturbation to otherwise
  constant atmosphere parameters (termed ``{\em Gaussian inversion}'' in the following). The perturbation is specified by its
location and its width in optical depth, which are identical in all atmospheric quantities, and a variable amplitude of
the perturbation in each parameter. However, still two magnetic components are
used: one component has constant properties with depth, and the second
component consists of the latter plus the added perturbation.  This allows
  to model that the LOS may encounter an uncombed topology only in a part of
  the pixel, which could well happen at the spatial resolution of the
  observations. The respective fill factor of both components is also a free
parameter in the inversion. Figure \ref{gausmodel} shows how the two
components of the Gaussian inversion look like in the best-fit solution of one
set of penumbral Stokes profiles. The total number of free parameters is 8 for
the constant background field component (field strength $B$, inclination
$\gamma$, azimuth $\phi$, LOS velocity $v$, 3 nodes for the temperature
stratification $T$, a constant micro-turbulent velocity $v_{mic}$); the
Gaussian perturbation adds again 8 more free parameters (amplitude of Gaussian
in $B, \gamma, \phi, v, T, v_{mic}$, location $\tau_{center}$, width
$\sigma$). The additional parameters shared by both components are an
identical macro-turbulent velocity $v_{mac}$, the fill factor of each component
$f$, and the stray light contribution to the spectra, $\beta$. This gives in
total 19 free parameters for 329 wavelength points $\times$ 4 Stokes
parameters (= 1316 measurement values), which, however, are not all fully
independent \citep[see the discussion in][]{almeida2005}. SIRGAUS has
  been used in a number of investigations
  \citep{bellot2003,beckthesis2006,jurcak+etal2007,cabrerasolana+etal2008,ishikawa+etal2010}. The main difference between the standard version of SIR and SIRGAUS is that spectral lines cannot be treated as blends of each other, i.e., the mutual influence of the line wings on the neighboring lines is not included. For that reason, also the two weakest lines in the observed spectra (\ion{Fe}{i} at 1564.74\,nm and \ion{Fe}{i} at 630.35\,nm) were not used for the fit since they are located in the wings of stronger lines.

I found that the Gaussian inversion is very sensitive to the initial model atmospheres, i.e., the code  fails to converge if the observed profiles are too far off from those resulting from the initial model. To
overcome this difficulty, I used the results of the 2C inversion as
input. The initial amplitude of the Gaussian was derived by subtracting the best-fit values of the atmospheric parameters of the two components in the 2C inversion. This introduces some bias in the inversion results, since the 2C inversion already gives a reasonable fit to the observed spectra \citep[cf.][]{beck2008}. However, the main properties of the field components like field strength and field inclination are ``robust'' quantities when using the near-IR spectral lines. These lines give rather strict limitations on for instance the field strength due to their strong splitting \citep[][Fig.~7]{beck+etal2006d}. The Gaussian inversion also is aimed at retrieving the vertical organization of the field components rather than their basic properties. The initial location and width of the Gaussian perturbation were set to $\log \tau = -0.5$ and 0.5 units of $\log\tau$, respectively. The same initial values were used for every inverted pixel in the penumbra, since these are exactly the quantities that the inversion should determine. 

Appendix \ref{inicomp} shows a comparison between the initial model atmospheres as derived from the 2C inversion and the final best-fit solutions of the Gaussian. The basic properties of the magnetic vector field (field strength and orientation) changed only slightly. Figure \ref{figbdiff} shows that the field topology corresponds to the predictions for an uncombed model with horizontal flow channels: the difference of field inclination between the two components, $\Delta\gamma$, is positive throughout the spot, whereas the difference of field azimuth $\Delta\Phi$ changes sign across the symmetry line of the spot \citep[cp.~][their Fig.~11]{mueller+etal2002}.
 
\begin{figure*}
\hspace*{.5cm}\resizebox{8cm}{!}{\includegraphics{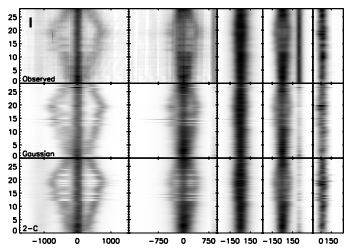}}\hspace*{.5cm}
\resizebox{8cm}{!}{\includegraphics{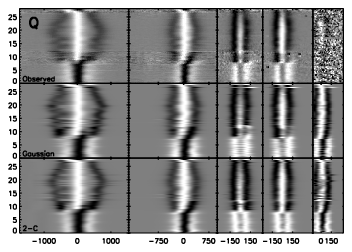}}\\$ $\\
\hspace*{.5cm}\resizebox{8cm}{!}{\includegraphics{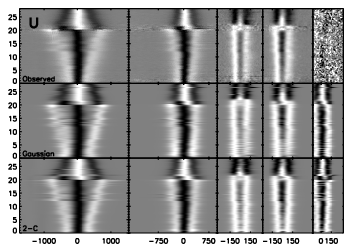}}\hspace*{.5cm}
\resizebox{8cm}{!}{\includegraphics{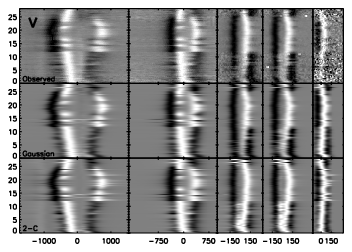}}\vspace*{.2cm}
\caption{Spectra of a spatial cut along the neutral line of Stokes $V$. {\em Clockwise, starting left top}: Stokes $IQVU$. Each subpanel shows the observed spectra in the {\em top row}, the best-fit profiles of the Gaussian inversion in the {\em middle row}, and those of the 2C inversion in the {\em bottom row}. {\em Left to right in each subpanel}: 1564.8\,nm, 1565.2\,nm, 630.15\,nm, 630.25\,nm, 630.37\,nm. The x-axis (dispersion) is in m{\AA}, the y-axis (spatial position along the cut) in arcsec. All profiles were normalized individually to improve the visibility of the spectral patterns; the range is $\pm 1$.\label{neutcut}}
\end{figure*}
\section{Results\label{sect_results}}
\subsection{Individual profiles and fit quality}
Figure \ref{neutcut} shows the observed and best-fit profiles of both the 2C
and the Gaussian inversion on a cut along the neutral line of Stokes $V$
(marked in Fig.~\ref{fig1}) for a visual control of the quality of the
inversion. Each row corresponds to the spectrum of a single pixel. The spectra
have been normalized separately to their respective maximal value in $IQUV$ to
improve the visibility of spectral features. The polarization signal of the
\ion{Ti}{i} line at 630.37\,nm is close to the noise level outside the umbra,
it only shows up clearly in Stokes $V$ at the locations where the splitting of
for instance 1568.4\,nm is largest ($y\sim20^{\prime\prime}$). The complex
shape of the profiles can be seen in the Stokes $V$ graph ({\em lower
  right}). The $V$ signals of all lines often show a local minimum {\em and} a
maximum in the blue lobe ({\em black/white}), and in the red lobe either a
minimum {\em or} a maximum \citep[see also][]{franz+schlichenmaier2010}. The
cut actually crosses the neutral line of the sunspot, the main polarity of the
Stokes $V$ signal changes at around $y\sim12^{\prime\prime}$. Both inversion
setups reproduce the observed Stokes spectra fairly well; neither differences
between observed and best-fit profiles nor between the two sets of best-fit
profiles can be easily discerned. One case of a clear difference can be found
in the Stokes $V$ signals of the VIS lines at 630.15\,nm and 630.25\,nm ({\em
  3rd and 4th column in the lower right panel}). In the lower part from
$y=0^{\prime\prime}$ to $12^{\prime\prime}$, the observed $V$ signals show a
(weak) minimum ({\em black}) followed by a broad maximum ({\em grey to white})
in the blue lobe, and a broad minimum ({\em black}) in the red lobe. The 2C
inversion is unable to clearly reproduce the broad minimum in the red
lobe. Another deviation between observations and best-fit profiles is seen in
Stokes $Q$ of again the VIS lines: in the same lower part of the spectra, the
Stokes $Q$ of the Gaussian inversion shows strong pixel-to-pixel variations
in, e.g., 630.15\,nm with Doppler excursion to the blue and to the red that are
absent in the observations. Appendix \ref{prof_examples} contains six
  examples of individual observed profiles, with the best-fit profiles of
both inversion setups overplotted. The location of the profiles is marked
  in Fig.~\ref{fig1}. 

To quantify the differences between the two inversion setups, I calculated the $\chi^2$ values of the squared difference between observed ($S_{j,obs}$) and best-fit profiles ($S_{j,synth}$) for both inversion setups by
\begin{equation}
\chi^2_{ij} = \sum_{\lambda_i} (S_{j,obs} - S_{j,synth})^2(\lambda_i) \;,
\end{equation}
where $i$ cycles through the spectral lines and $j$ through the entries of the Stokes vector, $S$.

\begin{figure}
\centerline{\resizebox{8.8cm}{!}{\hspace*{.3cm}\includegraphics{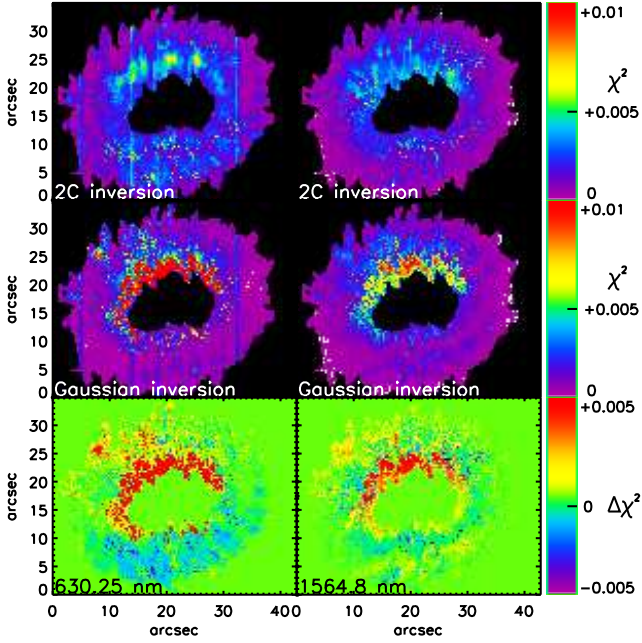}}}$ $\\
\caption{$\chi^2_V$ of the two inversion methods. {\em Top row}: $\chi^2_V$ of 630.25\,nm in the 2-component inversion ({\em left}), same for 1564.8\,nm ({\em right}). {\em Middle row}: same as above for the Gaussian inversion. {\em Bottom row}: difference of $\chi^2_V$ between the Gaussian and 2C inversion. Negative values indicate a better fit quality of the Gaussian inversion.\label{chidiff}}
\end{figure}

The {\em bottom row} of Fig.~\ref{chidiff} shows the difference of $\chi^2_V$ between the 2C inversion and the Gaussian inversion for the spectral lines at 630.25\,nm and at 1564.8\,nm. {\em Blue color} indicates a smaller $\chi^2$ in the Gaussian inversion (=improved fit quality), {\em red color} a larger value. For 630.25\,nm, the Gaussian inversion improved the fit quality on the limb side of the spot ({\em lower half}), whereas for 1564.8\,nm there are no visible changes there. On the center side in the innermost penumbra, $\chi^2_V$ got {\em worse} in the Gaussian inversion ({\em upper half}). The same profiles in the inner center side penumbra already show a worse fit in the 2C inversion as compared with profiles on the limb side. The reason is that the profiles on the limb side show clear signatures of two different magnetic field components (parallel and anti-parallel to the LOS) that contribute to the observed spectra, especially in the $V$ spectra from near the neutral line with more than two lobes \citep[e.g.,][]{grigorjev+katz1972,schlichenmaier+collados2002,almeida2005}. The profile shape of the near-IR lines varies even stronger than that of the 630\,nm lines \citep[see Figs.~\ref{prof1} to \ref{prof3}, or ][]{ruedi+etal1999,iniesta+etal2001,schlichenmaier+collados2002,beck2008}. On the limb side, this signature gets lost due to the fact that the field lines of both magnetic components are parallel to the LOS. The spectra there seemingly do not provide enough information to fix the free parameters in the Gaussian inversion. \citet{deltoroiniesta+etal2010} in a recent publication suggest to use Occam's razor principle for such a case: reject the more complex solution (the Gaussian inversion) in exchange for the simpler one (2C inversion). This approach, however, is not necessarily justified even in the case that the more complex solution gives a {\em worse} $\chi^2$. The 2C inversion has intrinsically a zero NCP, whereas the Gaussian inversion actually still reproduces the observed NCP to some degree, even at the cost of a worse $\chi^2$. The more complex solution therefore should be preferred over the simpler one for being a more realistic model approach for the presumably ``real'' solar magnetic field topology.
\subsection{Net Circular Polarization}
The NCP was defined by
\begin{equation}
{\rm NCP} = \int_{\lambda_0}^{\lambda_1} \frac{V}{I}(\lambda) d\lambda \,. \label{eqncp}
\end{equation}
Note that $V(\lambda)$ is not normalized with the continuum intensity, $I_c$,
but the intensity at the corresponding wavelength, $I(\lambda)$. Especially
for the deep VIS lines the normalization by $I_c$ can change the NCP value
significantly. I applied Eq.~(\ref{eqncp}) to all spectral lines in the
observations and the best-fit profiles separately, restricting the integration
range [$\lambda_0,\lambda_1$] to encompass only a single line each time. I
remark that the NCP is {\em no} quantity, whose deviation between observed and
synthetic spectra is minimized in the fit procedure. The SIRGAUS code minimizes the total deviation, $\chi_{tot}^2 = \sum_{ij} \chi_{ij}^2$, between the observed and the synthetic profiles. If an actual misfit of the NCP value reduces $\chi_{tot}^2$, the code is nonetheless forced to choose the solution with the ``worse'' NCP. 
With Eq.~(\ref{eqncp}), the derivation of the NCP is straightforward, but I think that a word of caution is appropriate. The NCP is a favorite toy for theoretical considerations, but from an observational point of view it is an ill-determined quantity. The NCP corresponds to a subtraction of, even with the integration in wavelength, two ``small'' numbers, the areas below the circular polarization lobes. Due to the low polarization signal level of at maximum about 40 \% of $I$, the NCP value is severely affected by the noise level in observations. Thus, even if the NCP is one of the few quantities directly dependent on the vertical structure of the solar atmosphere, its absolute value has to be treated with care whenever observations are concerned. 
\begin{figure}
\center
\resizebox{8.cm}{!}{\hspace*{.5cm}
\begin{minipage}{4cm}
\resizebox{4cm}{!}{\includegraphics{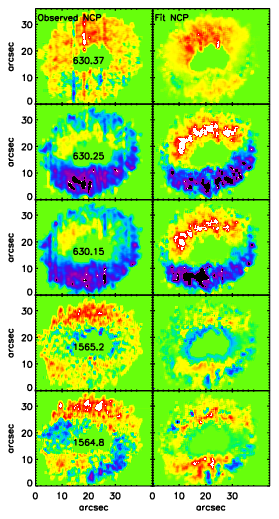}}$ $\\
\end{minipage}
\begin{minipage}{1.cm}
\resizebox{.95cm}{!}{\includegraphics{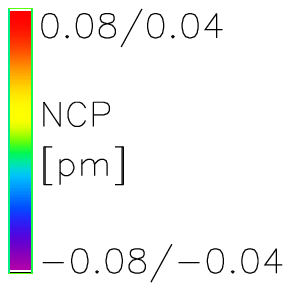}}\vspace*{.02cm}\\
\resizebox{.95cm}{!}{\includegraphics{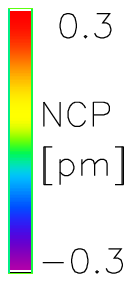}}\vspace*{.02cm}\\
\resizebox{.95cm}{!}{\includegraphics{zbarncp.ps}}\vspace*{.02cm}\\
\resizebox{.95cm}{!}{\includegraphics{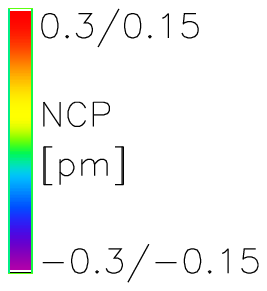}}\vspace*{.02cm}\\
\resizebox{.95cm}{!}{\includegraphics{zbarncp.ps}}$ $\\
\end{minipage}}
\caption{Comparison between the NCP in the observed ({\em left column}) and best-fit profiles ({\em right}). {\em Top to bottom}: 630.37\,nm, 630.25\,nm, 630.15\,nm, 1565.2\,nm, 1564.8\,nm. For the fit result of 630.37\,nm and of 1565.2\,nm, the display range is half that of the observed NCP.\label{ncpcomp}}
\end{figure}

Figure \ref{ncpcomp} shows a direct comparison of observed and best-fit NCP
for the whole sunspot. The display range for the best-fit NCP had to be halved
for 1565.2\,nm and 630.37\,nm to allow seeing the spatial variation across the
spot at all. The (anti-)symmetry properties of the NCP are recovered
completely only for the VIS lines, with positive (negative) NCP values on the
(center) limb side. The best reproduction of the large-scale spatial pattern
of the NCP is found for 630.37\,nm. For 630.15\,nm and 630.25\,nm, the center
side is less well reproduced. The negative NCP values at the outer penumbral
boundary in the observations are missing in the best-fit NCP, and the values
in the mid and inner center side penumbra are larger and strictly positive in
the best-fit NCP. For the IR lines, the agreement between observed and
best-fit NCP is not much better. The antisymmetric pattern of two minima and
maxima on an azimuthal path can be seen in the best-fit NCP for 1565.2\,nm and
1564.8\,nm, but the NCP amplitude is off by a factor of 2 for 1565.2\,nm like
for the \ion{Ti}{i} line. Again, the agreement on the center side is generally
worse than for the limb side. An interesting feature is seen near the
umbral-penumbra boundary. The umbra is encircled by a ring of negative
(positive) NCP in the 1565.2 (630.37)\,nm line in the observations, which does
not appear for the other spectral lines. The best-fit NCP actually reproduces
this feature, with opposite signs in the two spectral lines, even if this
includes exactly those locations where the $\chi_V^2$ of  1564.8\,nm and
630.25\,nm was worse.
\begin{figure*}
\centerline{\includegraphics{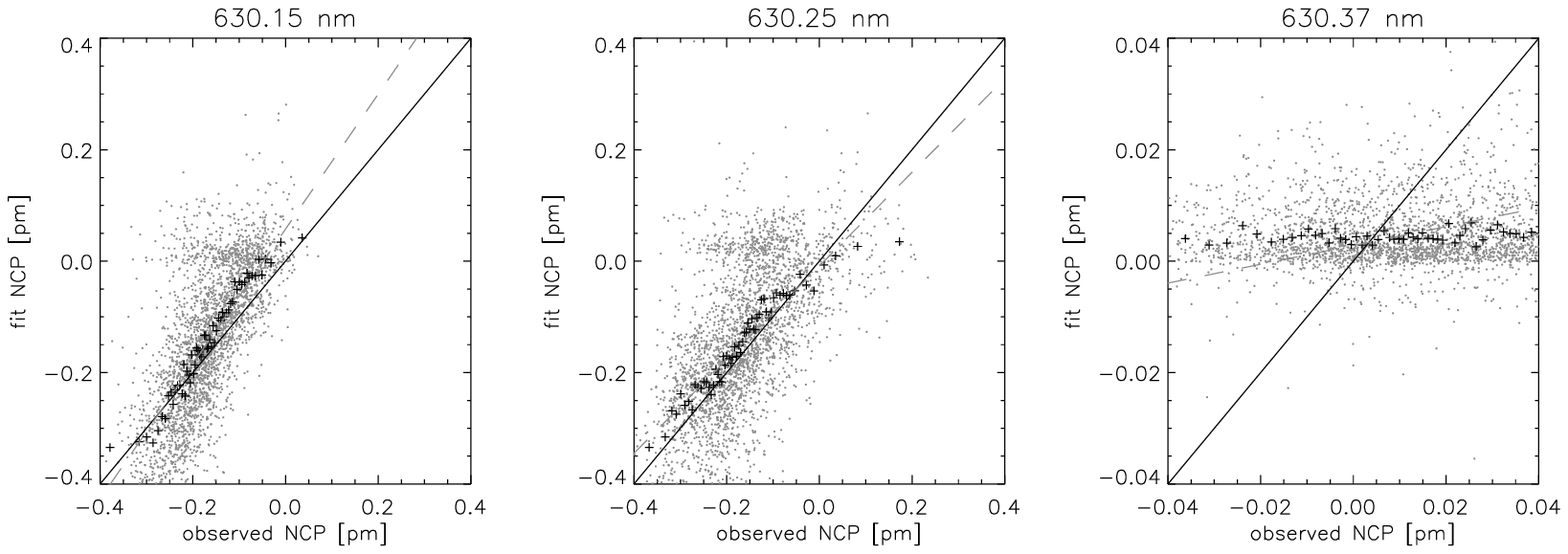}}
\centerline{\includegraphics{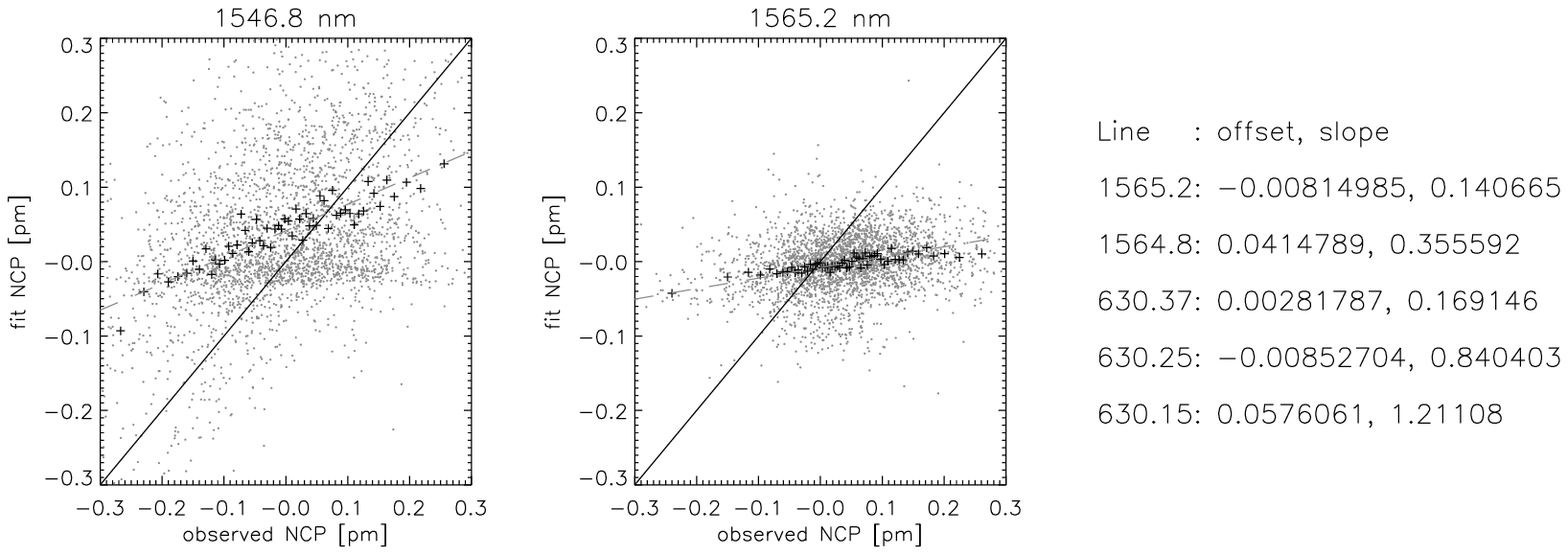}}
\caption{Scatterplots of NCP on the limb-side in the observed and best-fit spectra. {\em Grey dots} show all data points, {\em black pluses} the same with binning. The  {\em grey dashed} lines are linear regression lines with the results for offset and slope as given in the lower right. \label{scatncp}}
\end{figure*}

To investigate how far the inversion also catched the absolute value of the
NCP, I show scatterplots of observed vs.~best-fit NCP values for all spectral
lines (Fig.~\ref{scatncp}). I restricted the area considered to the limb side
of the spot since the 2-D maps of Fig.~\ref{ncpcomp} already suggest that no
clear correlation can be expected on the center side. The later detailed
investigation of the inversion results also will focus on the limb side. For
630.15\,nm and 630.25\,nm, the relation between observed and best-fit NCP is
close to linear with a slope of about unity. The best-fit NCP values are
slightly overestimating the observed NCP for 630.15\,nm (slope 1.21) and
underestimating it for 630.25\,nm (slope 0.84) \citep[cp.][Figs.~18 and 5,
respectively]{almeida2005,borrero+etal2006}. For both near-IR lines and
630.37\,nm, the fit NCP is significantly smaller than the observed NCP, with
only a weak trend between observed and best-fit NCP.
\begin{figure}
\resizebox{7cm}{!}{\includegraphics{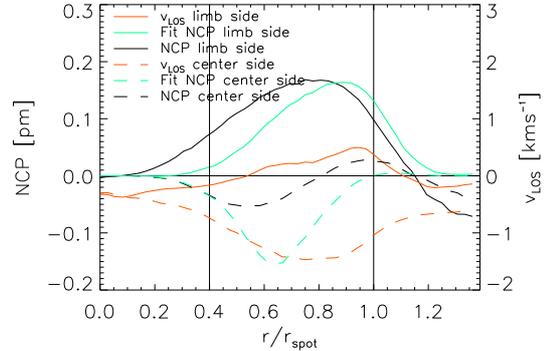}}
\caption{Radial variation of the observed NCP ({\em black}), best-fit NCP ({\em green}), and the LOS velocity ({\em orange}) on the limb side ({\em solid lines}) and the center side ({\em dashed lines}). The {\em vertical solid lines} denote the average inner and outer penumbral boundary. \label{radvar}}
\end{figure}
\subsection{Radial variation of NCP and $v_{LOS}$}
\citet[][T07]{tritschler+etal2007} pointed out the existence of a zero-crossing of the NCP of the 630.25\,nm line on a radial cut on the center-side, where the NCP changes sign in the mid to outer penumbra. The maps in Fig.~\ref{ncpcomp} support this finding, which also applies to the 630.15\,nm line as well. For a direct comparison with their Fig.~3, I determined the radial variation of the NCP and the LOS velocity for the 630.25\,nm line (cf.~Fig.~\ref{radvar}). The zero-crossing of the center side NCP is located at a similar radial position of around 0.8 $r/r_{\rm spot}$ as in T07, presumably to the similar heliocentric angles of the observations (30$^\circ$ and 42$^\circ$, respectively). The location of the zero-crossing is co-spatial to the maximum of the LOS velocity on the center-side ({\em orange dashed line}). Figure \ref{radvar} also demonstrates that the best-fit NCP of 630.25\,nm on the limb side follows closely the observed NCP qualitatively and, in this case, even quantitatively. 
\subsection{Topology of the flow channels from the location and the width of the Gaussian perturbation\label{subsec_flowch}}
\begin{figure}
\centerline{\includegraphics{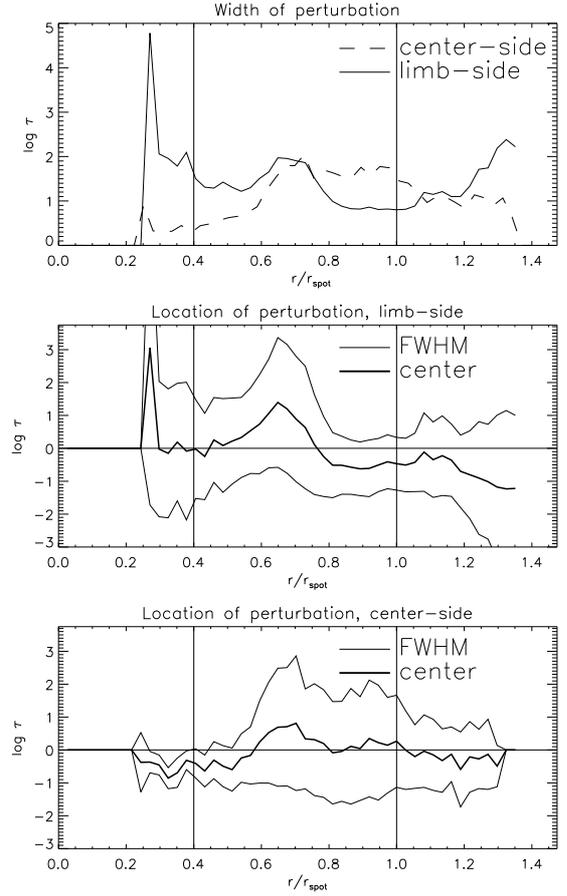}}
\caption{Parameters of the Gaussian perturbation. {\em Top}: Radial variation of the HWHM of the Gaussian
  perturbation in the inversion in units of log $\tau$. {\em Middle}: location of the center of the perturbation for the limb-side ({\em thick line}). The {\em thin lines} indicate the FWHM. {\em Bottom}: same as above for the center-side.\label{invresgaus}}
\end{figure}

The information on the location and spatial extent of the flow channels is contained in the central position $\tau_{center}$ and the width $\sigma$ of the Gaussian perturbation. In the following, I take the full width at half maximum (FWHM) of the perturbation as a measure of the vertical extent of the flow channels. Figure \ref{invresgaus} displays the azimuthally averaged values of location and width in optical depth units as function of the radial distance to the spot center, separately for the center and the limb side. For $r/r_{spot} <$ 0.6, the center of the perturbation always stays close to the log $\tau$ = 0 level. The width is so large that the lower boundary is close to or below the log $\tau$ = 0 level. Around  $r/r_{spot} =$ 0.65, there is a distinct maximum in the location value. The maximum is more pronounced in the limb-side. The half-width-at-half-maximum (HWHM, {\em top panel}) decreases from about 2 to 1 units of optical depth after the maximum in the location. For $r/r_{spot} >$ 0.8, the location is roughly constant at log $\tau =$ -0.5 on the limb side and  at log $\tau =$ 0 on the center side, respectively.

For an interpretation of the behavior one has to take into account that the
values of both parameters are given in the optical depth scale. If the Gaussian perturbation is located at log $\tau$ = 0, it can be interpreted in two ways: it
may indicate a low-lying flow channel {\em or} indicate that it is an optically
thick structure. Thus, the isolated maximum at $r/r_{spot} =$ 0.65 can either indicate flow channels sinking (location in log$\tau$ changes from 0 to +1.5) and immediately again rising (location in log$\tau$ changes from +1.5 to 0) in the mid penumbra, which seems improbable, or that they are optically thick up to a radius of $r/r_{spot} =$ 0.65. The following decrease of the location from log$\tau$ = 1.5 to around -0.5 with increasing radial distance suggests in my opinion that the latter interpretation is true. As long as the LOS can not penetrate down to the lower boundary of the flow channel, the inversion code has no information on where to place it. The decrease of the location for larger radii could then be interpreted as the rise of a flow channel through the surface, where the
profiles then contain the information necessary to place it in the
atmosphere. Another argument for this interpretation are the relative fill
factors of the two inversion components. At $r/r_{\rm spot} =$ 0.6, the fill factor of the stronger inclined component identified with the flow channels strongly increases \citep{bellot+etal2004,beck2008}. This indicates that the flow channels start to have a stronger contribution to the observed profiles which could be due to the fact that they breach the surface there. It also indicates that for the inner penumbra the location of the Gaussian perturbation derived by the inversion code may be unreliable. In the inner penumbra, the Gaussian inversion reproduces the NCP with the discontinuity at the upper boundary of the flow channel, but it cannot locate it in optical depth because the LOS does not penetrate through the flow channel. 

It is worthwhile to compare Figs.~\ref{invresgaus} with the corresponding Fig.~6 of \citet{borrero+etal2006}. The flow channels in the current investigation are globally located at lower optical depths, which could be the result of the initial atmosphere stratification used for the background component (penumbral model of \citet{toroiniesta+etal1994} vs.~HSRA model, respectively). Otherwise, the radial variation follows similar trends: the location is near log $\tau$ = 0 in the innermost penumbra, moves upwards in optical depth at around 0.7 $r/r_{spot}$ and shows a slight decrease at (beyond) the outer sunspot boundary. 
\begin{figure}
\centerline{\resizebox{7cm}{!}{\includegraphics{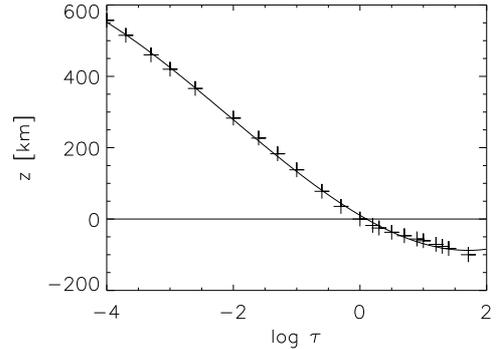}}}
\caption{Conversion curve between optical depth, $\tau$ and geometrical height, $z$, from the HSRA model. +: HSRA, {\em solid line}: polynomial fit of 3th order.\label{tautogeo}}
\end{figure}
\paragraph{Conversion to geometrical height} I converted the optical depth scale to a geometrical height scale using the tabulated values of the Harvard Smithsonian Reference Atmosphere (HSRA) model \citep{gingerich+etal1971} for the conversion from $\tau$ into km above $\tau_{500} = 1$ (cf.~Fig.~\ref{tautogeo}). Even if the HRSA model atmosphere has been derived as an average stratification of the quiet Sun devoid of magnetic fields, it should still be usable as first approximation. I assumed additionally that the iso-surface of $\tau_{500} = 1$ has a slope of 3 deg when going from the inner penumbral boundary towards the outer boundary, corresponding to a Wilson depression of 380 km in the umbra \citep{schliche+schmidt2000}. Recently, \citet{puschmann+etal2010} have presented a more solid determination of a geometrical height scale for inversions of spectropolarimetric data that eventually could be employed to the present results as well. Figure \ref{fincomp1} shows how the location and width in optical depth appears in geometrical height with the assumptions above. For a comparison, I overplotted arrows that give the field inclination of the flow channel component in the 2C inversion. The radial variation of the location of the Gaussian perturbation agrees quite well with the field inclination from the 2C inversion.

The results on the location of the perturbation seem to be unreliable for $r/r_{spot} <$0.65 ({\em vertical black line} in Fig.~\ref{fincomp1}), because of the counterintuitive descend and rise of the location of the perturbation around that radius (cf.~Fig.~\ref{invresgaus}). For $r/r_{spot} <$0.65, I thus substituted the values derived from radially integrating the field inclination to the surface like described in \citet{beck2008}. The location of the flow channel left of the {\em black vertical line} thus is based on the 2C inversion results. The final result of combining the 2C and Gaussian inversion for the whole penumbra then is a flow channel that ascends steeply in the inner penumbra, turns into a slightly elevated flow channel which is close to horizontal in the outer penumbra, and slightly bends down in the outermost penumbra.
\begin{figure}
\centerline{\resizebox{8.8cm}{!}{\includegraphics{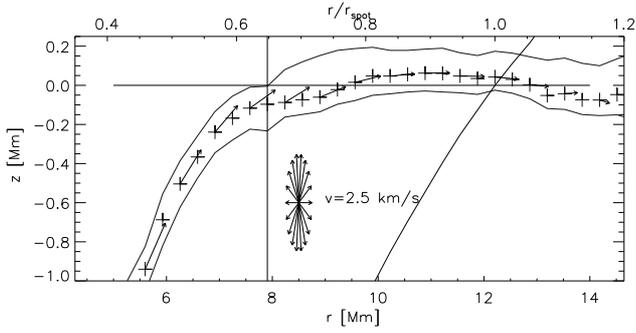}}}
\caption{Topology of the average flow channel from the Gaussian inversion
    in geometrical height (limb side only). The {\em inclined black line} starting at r $\sim$ 10 Mm gives the integrated inclination of the background field. The direction of the arrows gives the field inclination in the 2C inversion, their length is proportional to the velocity. {\em Crosses} mark the location of the center of the flow channel.\label{fincomp1}}
\end{figure}

The conversion to the geometrical height scale also allows to derive the diameter of the flow channel in km. The non-linear relation between optical depth and geometrical height (Fig.~\ref{tautogeo}) in combination with a FWHM given in units of log$\tau$ by the inversion makes the flow channel asymmetric around the central location of the perturbation in the geometrical height scale. I then defined the FWHM in km as the difference between the location of the upper and lower boundary that resulted from transforming $\tau_{center}\pm$HWHM (in units of log$\tau$) to the geometrical scale. The inversion code determined the FWHM along the LOS which can cut the assumed flow channel at some angle (see Fig.~\ref{flux_calc}). The FWHM should thus be corrected for the field inclination to the LOS by a multiplication with $\cos\,( |90^\circ - \gamma_{LOS, fc}|)$. Since $\gamma_{LOS, fc}$ is close to 90\,deg on the limb side, the effect is, however, small ({\em dotted line} in Fig.~\ref{widthgeo}); I have thus not re-drawn Fig.~\ref{fincomp1} with the corrected FWHM. Within all the limitations imposed by the analysis and the conversion to geometrical height, the results indicate a rather constant diameter of the flow channel with a FWHM of about 250 km (cf.~Fig.~\ref{widthgeo}).
\subsection{Magnetic flux of the two components}
The relative amount of magnetic flux in the two magnetic components
representing the flow channels and the background field is of interest for the
question if the penumbral heating can be achieved by the repetitive rise or
motion of hot flow channels. The calculation of the magnetic flux from the
inversion results is, however, not straightforward. Figure \ref{flux_calc}
visualizes the way the inversion code has constructed its synthetic spectra:
the 3-D volume element given by the spatial extent of the pixel $(\Delta x,
\Delta y)$ = (slit width, spatial sampling along the slit) and the optical
depth axis is first separated in two different regions by the fill factor of
the magnetic components, $f$. The fill factor is defined in the plane
perpendicular to the LOS. Two different atmospheric stratifications along the
LOS are then assumed in the two regions: in one case only the constant
background field, in the other case the bg field plus the Gaussian
perturbation. For the second component with the Gaussian perturbation, the
plotted cylinder is actually not fully correct, since the inversion code used
in principle only a Gaussian shape along the optical depth axis, not in the
spatial dimension. Both bg field and fc field can have an arbitrary orientation to the LOS and to each other; I only sketched three cases with a different field inclination of the fc component. The magnetic field inclination is available for both components in the LOS reference frame or relative to the local surface normal; I will use the inclination to the surface normal in the following. For the background field, I then determined the magnetic flux as
\begin{figure}
\centerline{\includegraphics{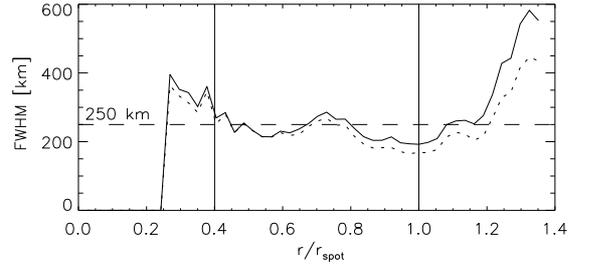}}
\caption{FWHM of the average flow channel on the limb side ({\em solid}). The {\em dotted line} shows the FWMH after correction for the field inclination to the LOS. The {\em dashed vertical line} denotes a FWHM of 250\,km.\label{widthgeo}}
\end{figure}
\begin{equation}
\Phi_{bg}(r) = (B_{bg}\cdot \cos \gamma_{bg} \cdot f_{bg})(r) \cdot A_{res} \;,
\end{equation}
where $A_{res} = (0.36 \cdot 725)^2 {\rm km}^2$ is the area corresponding to a pixel, $\gamma_{bg}$ the magnetic field inclination relative to the surface normal, and $r$ measures the distance to the spot center. The fill factor $f_{bg}$ was set to 1, since the background field is present in both inversion components with identical properties.

For the flow channel component, I first converted the fill factor to an effective size $d_{eff}$  in the x-y-plane. Since the inversion cannot provide information about the spatial organization of the two components but only the total fill factor inside the pixel, the definition of $d_{eff}$ is ambiguous. I used two different solutions for $d_{eff}$. In the first, I defined $d_{eff 1}$ as the edge length of a square whose area is equal to the fill factor by
\begin{equation}
d_{eff 1} = \sqrt{ A_{res} \cdot f_{fc} } \;,
\end{equation}
like sketched in the {\em upper right} graph of Fig.~\ref{flux_calc}. This solution makes use of the fact that the two spatial dimensions $(x,y)$ of the pixel have identical sizes of 0\farcs36, thus a square shape is suggested. Taking into account the inclination of the fc component close to 90\,deg (either relative to the LOS or to the surface normal), this solution, however, provides a inconsistent topology since the nearly horizontal field lines of the fc component would have to terminate abruptly at some spatial location. A second solution is to demand that one of the axes $(x,y)$ of the area that corresponds to the fill factor has to extend along the full size of the 3-D volume ({\em bottom graph}). Then the effective size in the plane perpendicular to the orientation of the magnetic field lines is given by
\begin{equation}
d_{eff 2} = \frac{A_{res} \cdot f_{fc}}{261 {\rm km}} \;.
\end{equation}

Along the LOS, the inversion provided the stratification of the magnetic field with optical depth. The conversion to geometrical height in the previous section then yielded the FWHM of the Gaussian perturbation in km. I integrated the resulting Gaussian function
\begin{equation}
B_{fc}(r,z) = \exp\left( -\frac{ (z - z_0)^2}{2 \cdot \sigma^2(r)}\right) \cdot B_{fc}(r)
\end{equation}
over a  height range of 1000\,km in $z$, to cover the full extent of the fc component. The width $\sigma(r)$ was once derived from the FWHM values as shown in Fig.~\ref{widthgeo}, corrected for the inclination of the fc component to the LOS, and once set to correspond to a fixed FWHM of 250\,km.
\begin{figure}
\centerline{\resizebox{6cm}{!}{\includegraphics{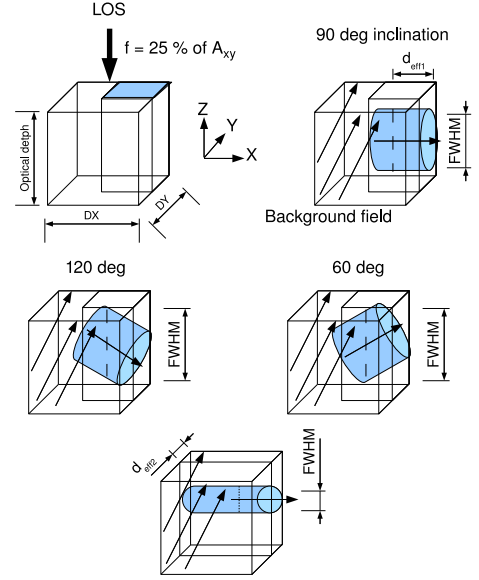}}}
\caption{Determination of the flow channels' area from the inversion results. {\em Left top}: definition of fill factor $f$. {\em Top right}: flow channel at 90\,deg to the LOS. The {\em dashed vertical line} denotes the FWHM determined by the code. $d_{eff 1}$ denotes the effective size of the fc component. The {\em parallel inclined arrows} denote the orientation of the bg field. {\em Middle row}: same for LOS inclinations of the fc component of 60 and 120\,deg. {\em Bottom}: another possible solution for the effective size $d_{eff 2}$.\label{flux_calc}}
\end{figure}

The total flux $\Phi_{fc}$ of the fc components was then derived from
\begin{equation}
\Phi_{fc}(r) =  d_{eff}(r) \cdot \int_{0 {\rm km}}^{1000 {\rm km}} B_{fc}(r,z)\, dz 
\end{equation}
for three cases, using the FWHM from the inversion and the two different definitions of $d_{eff}$ as given above, and once with a fixed FWHM and $d_{eff 1}$. The correction of the FWHM with the LOS inclination should provide that the effective area used is perpendicular to the field lines of the fc component. For the calculations of magnetic flux, I ignored the off-center position of the sunspot at about 30\,deg heliocentric angle. The assumed size $A_{res}$ as given by slit width and spatial sampling along the slit actually corresponds to a slightly larger area (1/cos$\, 30^\circ$ = 1.15) on the solar surface, but the projection effects enters both in the bg and fc flux and should thus be negligible.

\begin{figure}
\includegraphics{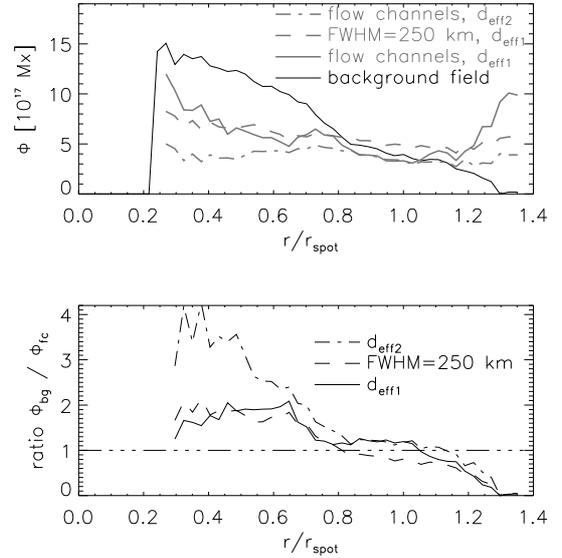}
\caption{Magnetic flux in the two inversion components. {\em Top}: total
  flux contained in the background field and flow channel (fc) component. {\em Bottom}: ratio of background field to flow channel component flux. The {\em horizontal line} marks a ratio of unity.\label{fluxcomp}}
\end{figure}

Figure \ref{fluxcomp} then displays the magnetic fluxes derived from the different approaches ({\em upper panel}), and the ratio between the flux of the bg and fc component ({\em lower panel}). Using the definition of $d_{eff 2}$ yields a flux $\Phi_{fc}$ of about 3.8$\times 10^{17}$ Mx per pixel that is almost constant throughout the spot, whereas $d_{eff 1}$ leads to a radial decrease of $\Phi_{fc}$. The flux of the bg component decreases with radius, due to the radial decrease of field strength and the increase of field inclination (see Fig.~\ref{velcomp2} below). Assuming that the the use of $d_{eff 2}$ provides the more consistent value for $\Phi_{fc}$, the ratio $\Phi_{bg}$/$\Phi_{fc}$ changes from about 4 at the umbral-penumbral boundary to 1 at the outer white-light boundary of the sunspot. This would imply that 20\% to 50\% of the magnetic flux in a sunspot participate in the Evershed flow and thus presumably also in the vertical energy transport throughout the penumbra.
\section{Discussion\label{sect_disc}}
I applied the uncombed ``Gaussian'' inversion with its capability to reproduce the NCP to the full penumbra of a sunspot, with the information of five spectral lines as input. The initial model of the fit was taken partly from a 2C inversion with constant magnetic field parameters to improve the convergence ($\Delta B, \Delta v, \Delta \Phi$, etc.), but no a priori information on the vertical structure was provided. The resulting best-fit spectra reproduce the observed NCP satisfactorily on the limb side for the VIS lines near 630\,nm and the \ion{Fe}{i} line at 1564.8\,nm, but fail partly on the center side. The NCP of the near-IR lines is also generally reproduced worse than that of the VIS lines, mainly in the amplitude of the best-fit NCP values that fall short of the observed values. There are some possible reasons for that, with the most important of course that in the inversion process the total least-squares deviation between observed and synthetic profiles is minimized, not the difference of the NCP, i.e., the mismatch in the NCP can be insignificant for the best-fit solution. The NCP of the near-IR lines also depends stronger on the exact location and the width of the Gaussian perturbation since their formation height is smaller than for the VIS lines \citep{cabrera+bellot+iniesta2005}. The Gaussian shape of the perturbation sets an upper limit on the steepness of the gradients in velocity and magnetic field; it does not correspond to a sharp discontinuity as for instance in \citet{borrero+etal2007}. The visible lines with their larger formation height are less sensitive to the exact location of the perturbation, and the gradients along the LOS contribute over a larger range in optical depth. The inversion code often used a very broad Gaussian perturbation on the inner center-side penumbra which then reduced the resulting NCP of the near-IR lines. The sign change of the NCP of the VIS lines in the radial direction on the center side is not reproduced by the inversion, but a ring of positive (negative) NCP just around the umbra in the 630.37\,nm line (1565.2\,nm line) is. 

\begin{figure}
\resizebox{7cm}{!}{\includegraphics{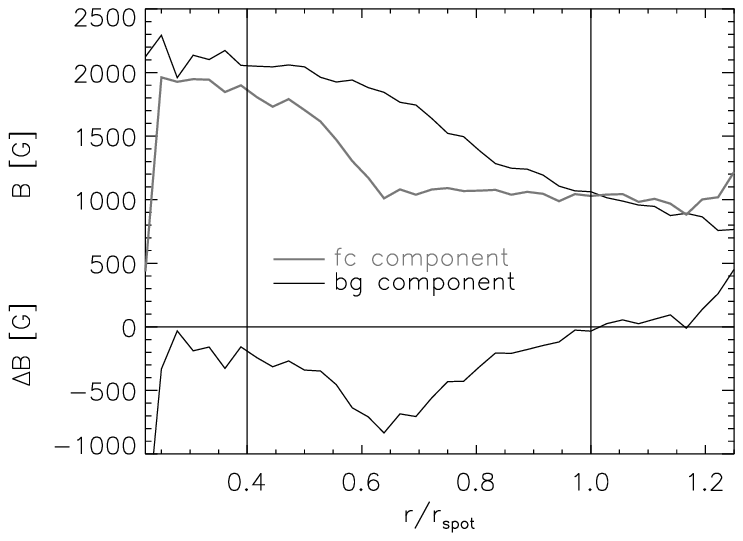}}\\
\resizebox{7cm}{!}{\includegraphics{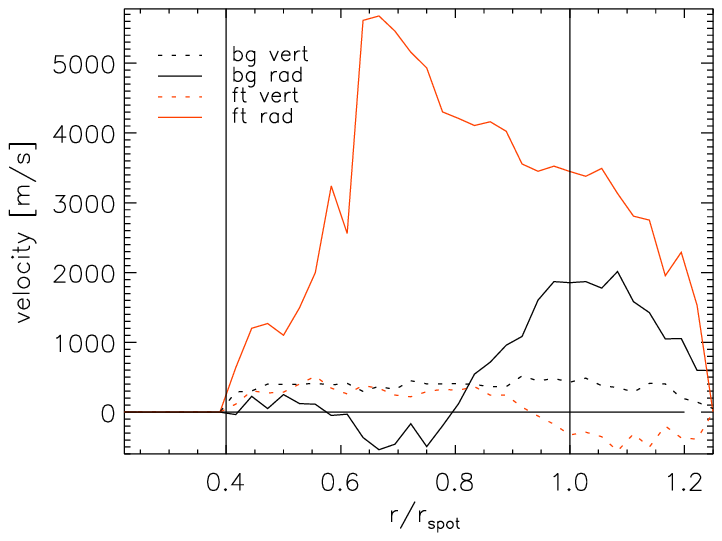}}
\caption{Radial variation of magnetic field strength and LOS velocity. {\em Top}: radial variation of the field strength in flow channel ({\em thick grey}) and background field ({\em thin black}). The difference between them is plotted in the lower half. {\em Bottom}: horizontal and vertical velocities for background field and flow channel component from the fit of a sinusoidal to the azimuthal variation of $\rm v_{LOS}$. {\em Black:} background field, {\em red:} flow channel component. {\em Dashed:} vertical velocity, {\em solid:} horizontal velocity.\label{velcomp2}}
\end{figure}

T07 suggested that the sign change of the NCP can be reproduced if the field strength in the flow channels is stronger by about 0.5 kG than in the background field. This proposed difference is in some conflict with the inversion results where the bg field is found to be stronger all throughout the penumbra, and of equal strength at the outer penumbral boundary \citep[][or Fig.~\ref{velcomp2}]{bellot+etal2004,borrero+etal2004,beck2008}. I suggest a different reason for the sign change, which also could explain why the Gaussian inversion was unable to reproduce even only the correct sign of the NCP in the center side. The bg component used parameters constant with optical depth, and thus, it did not contribute any NCP to the best-fit spectra. Without showing flow velocities as high as for the ``flow channels'', the bg component also has a significant velocity component of up to 2 kms$^{-1}$ in the outer penumbra \citep[{\em lower panel} of Fig.~\ref{velcomp2};][]{bellot+etal2004,borrero+etal2006}. The horizontal and vertical velocities have here been computed using the azimuthal variation of the LOS velocity all around the spot, also including the center side penumbra \citep[see e.g.][]{schliche+schmidt2000,beckthesis2006,beck2008}. Line-of-sight gradients in the flow velocity and additionally in the field strength of the bg component could create a non-zero NCP contribution from this component as well. Since the bg field lines are closer to being parallel to the LOS on the center side than on the limb side, this contribution would be more prominent in the Stokes $V$ profiles of the center side and could possibly produce the sign change of the NCP.

The conversion of the location of the Gaussian perturbation to a geometrical height scale was done with simplifying assumptions on the atmospheric density stratification. A common height scale as used in \citet{almeida2005} would be an option. I think, however, that a revised or improved conversion between optical depth and geometrical height will not change the picture of the flow channels' topology significantly. Changes of the location by for instance up to $\pm$100 km would, e.g.~not remove the flow channel from being close to the z=0 km or change its radial behavior strongly. The flow channel topology from the Gaussian inversion is in good agreement with the one derived in \citet{beck2008} from the integration of the field inclination, even if the two methods are fully independent of each other: the location of the perturbation is determined separately for each pixel in the inversion, whereas the integration uses the radial variation of the inclination to derive the geometry. 

The width of the flow channel as given by the inversion code is more critical. On several pixels, especially in the inner penumbra, the LOS cannot have penetrated through the Gaussian perturbation since the lower boundary is significantly below log $\tau$=0. The information on the vertical extent is missing in this cases, the code has only used the location of the upper boundary to produce the gradients needed to fit the spectral lines. In these location, the inversion results only provide information on the upper limit, a maximum height of about 200 km above the z=0 km level.

With the same caveat that the ``width'' of the fc structure seems an ill-defined quantity in some cases, the derived magnetic flux of the two components indicates that in the mid to outer penumbra about 20-50\% of the magnetic flux appears in the form of flow channels. This large fraction would imply that the temporal evolution of the penumbral fields has to be taken as the permanent and global re-arrangement of all magnetic field lines rather than isolated events of ascending flow channels. 

Throughout the analysis of the data presented here I now have used the
terminology of a ``background field'' and a ``flow channel'' component,
suggestive of an interpretation in terms of horizontal flux tubes embedded in
a less inclined field. If this picture is representative of the penumbral
magnetic fields is, however, an open question. At first, this terminology has to
be taken simply as a convenient way of describing the two distinct magnetic
components required to reproduce the observations, where one of them shows a
larger flow velocity, a weaker field strength, and a larger inclination to the
local surface normal than the other, similar to the ``minor'' and ``major''
component in \citet{almeida2005}. The two components can be combined into a
single one with strong LOS gradients \citep{mathew+etal2003,borrero+etal2004}
that still reproduces the observed spectra. This approach runs into the
problem that the magnetic field lines of sunspots cannot extend to the upper
solar atmosphere and the corona since one of the components requires nearly
horizontal fields, which is at odds with coronal observations. The analysis results thus require two distinct magnetic components in the penumbra, but they do not provide directly the possibility to choose between any of the three penumbral models (MISMA; field-free gaps; flux tubes). One argument in favor of the flux tube picture is the spatial radial coherence of penumbral filaments over several thousand km. In the MISMA picture, it is hard to envisage how such a large-scale structure can be formed by magnetic fields structured on the smallest scales without a common orientation. \citet{ichimoto+etal2007} recently reported the presence of roundish patches in a sunspot on disc center that appeared preferentially in the mid and outer penumbra and corresponded to strong downflows. \citet{almeida+ichimoto2009} interpreted this patches in the MISMA picture, but they also fit well to the flux tube model. In the outer penumbra, the inclination of the flow channel component shows on average a downward orientation \citep[e.g.,][]{beck2008}, which implies downflows for field-aligned mass motions. The revision of the original simulations of \citet{schliche+jahn+schmidt1998} in \citet{schlichenmaier2002} with the peculiar ``sea-serpent'' shape provides additional indications that also in the flux tube model downflows can be expected in the inner penumbra. The vertical velocity of the flow channel component in Fig.~\ref{velcomp2} shows the same: $v_{vert} < 0$\,ms$^{-1}$ ($\equiv v$ is oriented downwards in this case) for $r/r_{spot}>$ 0.9. A downward oriented flux tube as drawn in the {\em middle left panel} of Fig.~\ref{flux_calc} naturally would produce a roundish downflow patch in a horizontal cut. One caveat for the field-free gap model is also actually the failure of the Gaussian inversion to reproduce the sign change of the NCP on the limb side and the generally worse fit to the near-IR lines: the presence of gradients in LOS velocity and field strength is not sufficient for reproducing the observations \citep{scharmer+spruit2006}, only if they happen to be the correct gradients. 
\section{Conclusions\label{sect_concl}}
The uncombed ``Gaussian'' inversion is able to reproduce the NCP of simultaneous observations in VIS and near-IR spectral lines well on the limb side of a sunspot, where the signature of the two different magnetic components in the penumbra is clearest. Even with a mismatch in the NCP, it still is able to reproduce the observed spectra of five spectral lines satisfactorily all throughout the penumbra, with at least the same quality as a 2-component inversion with constant magnetic field properties (Figs.~\ref{prof1} to \ref{prof3}). The inversion setup can be interpreted as embedded flow channels like in the picture suggested by \citet{solanki+montavon1993}. As remarked by \citet{scharmer+spruit2006}, the agreement does not prove the correctness of the flux tube model, but it still proves that the model definitely is {\em not at odds} with the observations. 

Around 20-50\% of the total magnetic flux in the penumbra shows the characteristics of the component addressed as flow channel, implying a permanent re-organisation of all magnetic field lines in the penumbra. This could maybe allow to replenish the radiative losses of the penumbra also in the moving tube model of \citet{schliche+jahn+schmidt1998}. The calculations of \citet{schliche+bruls1999} yielded an energy supply by a single flux tube that was insufficient to compensate the penumbral energy losses on a radial cut, but eventually the rate and/or number of such tubes should be increased until their magnetic flux reaches 50\% of the static background field component.

Two peculiarities of the NCP observed in the sunspot at 30$^\circ$ heliocentric angle stand out prominently: a sign change of the NCP of the VIS lines on the center side as reported by \citet{tritschler+etal2007}, and a ring of positive and negative NCP just around the umbra in the \ion{Ti}{i} line at 630.37\,nm and in the \ion{Fe}{i} line at 1565.2\,nm, respectively. The first item is not reproduced by the inversion and is also missing in the theoretical calculations of, for instance, \citet{mueller+etal2002,mueller+etal2006}, or in \citet[][Fig.~19, for a sunspot closer to disc center]{almeida2005}. It would be interesting to see if this sign change is reproduced by the flux tube model used in \citet{borrero+etal2007} or \citet{borrero+solanki2010} on radial instead of azimuthal paths. The ring pattern around the umbra might be well suited for comparisons between observations and the MHD simulations of \citet{rempel+etal2009,rempel+etal2009a}, since the penumbral ``filaments'' in the simulations are somewhat shorter than any observed filaments, but provide information on the inner footpoints near the umbral-penumbral boundary.

The observations used in the present study were taken at the VTT in 2003, when it was only possible to improve the spatial resolution by a correlation tracker providing image stabilization, and with a 50\% loss of light due to the use of an achromatic beamsplitter to feed the two VIS and near-IR instruments. With the growing activity of the new solar cycle, it would now be possible again to obtain improved multi-wavelength data sets at the VTT for future studies, with both a higher signal-to-noise ratio thanks to a new dichroic BS and better spatial resolution thanks to adaptive optics.
\begin{acknowledgements}
The VTT is operated by the Kiepenheuer-Institut f\"ur Sonnenphysik (KIS) at the Spanish Observatorio del Teide, operated by the Instituto de Astrof{\'is}ica de Canarias (IAC). POLIS has been a joint development of the KIS and the High Altitude Observatory, Boulder, Colorado. C.B.~acknowledges partial support by the Spanish Ministry of Science and Innovation through project AYA 2007-63881. I thank L.R.~Bellot Rubio for his support during my thesis and fruitful discussions.
\end{acknowledgements}
\bibliographystyle{aa}
\bibliography{references_luis} 
\begin{appendix}
\section{Changes between initial and best-fit model atmosphere\label{inicomp}}
\begin{figure}
\center
\resizebox{5.9cm}{!}{\hspace*{.5cm}
\begin{minipage}{4cm}
\resizebox{4cm}{!}{\includegraphics{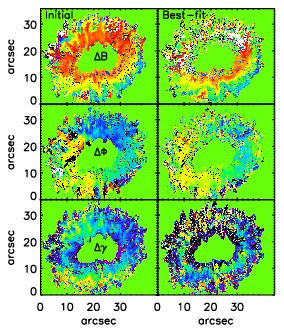}}
\end{minipage}
\begin{minipage}{1.cm}
\resizebox{.95cm}{!}{\includegraphics{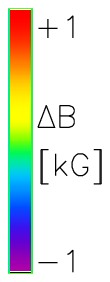}}\vspace*{.03cm}\\
\resizebox{.95cm}{!}{\includegraphics{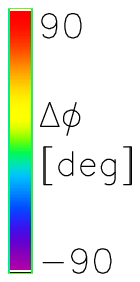}}\vspace*{.03cm}\\
\resizebox{.95cm}{!}{\includegraphics{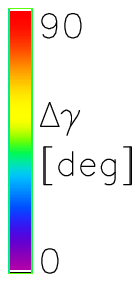}}
\end{minipage}}$ $\\$ $\\$ $\\
\caption{Difference of the magnetic field strength ({\em top row}), of the LOS field azimuth ({\em middle row}), and of the LOS field inclination ({\em bottom row}) between the two inversion components. {\em Left column}: initial model atmosphere. {\em Right column}: final best-fit model atmosphere, taken at log $\tau$ = 0.\label{figbdiff}}
\end{figure}

The Gaussian inversion was initialized with the difference between the two inversion components in the 2C inversion as amplitude of the Gaussian perturbation. Figure \ref{figbdiff} shows how much the initial value was modified in the inversion process. The changes in the difference of the LOS magnetic field azimuth, $\Delta\Phi$, the difference of the LOS magnetic field inclination, $\Delta\gamma$, or the difference in field strength, $\Delta B$, were actually minor. The field orientation ($\Delta\Phi$, $\Delta\gamma$) changed more than the field strength, but this could also be due to the fact that the values for the second component with the Gaussian perturbation were taken at a fixed optical depth of log $\tau = 0$.
\section{Profile examples\label{prof_examples}}
\begin{figure*}
\centerline{\bf\large 1 \hspace*{8.5cm} 2}$ $\\
\centerline{\resizebox{17.6cm}{!}{\includegraphics*[0,160][523,796]{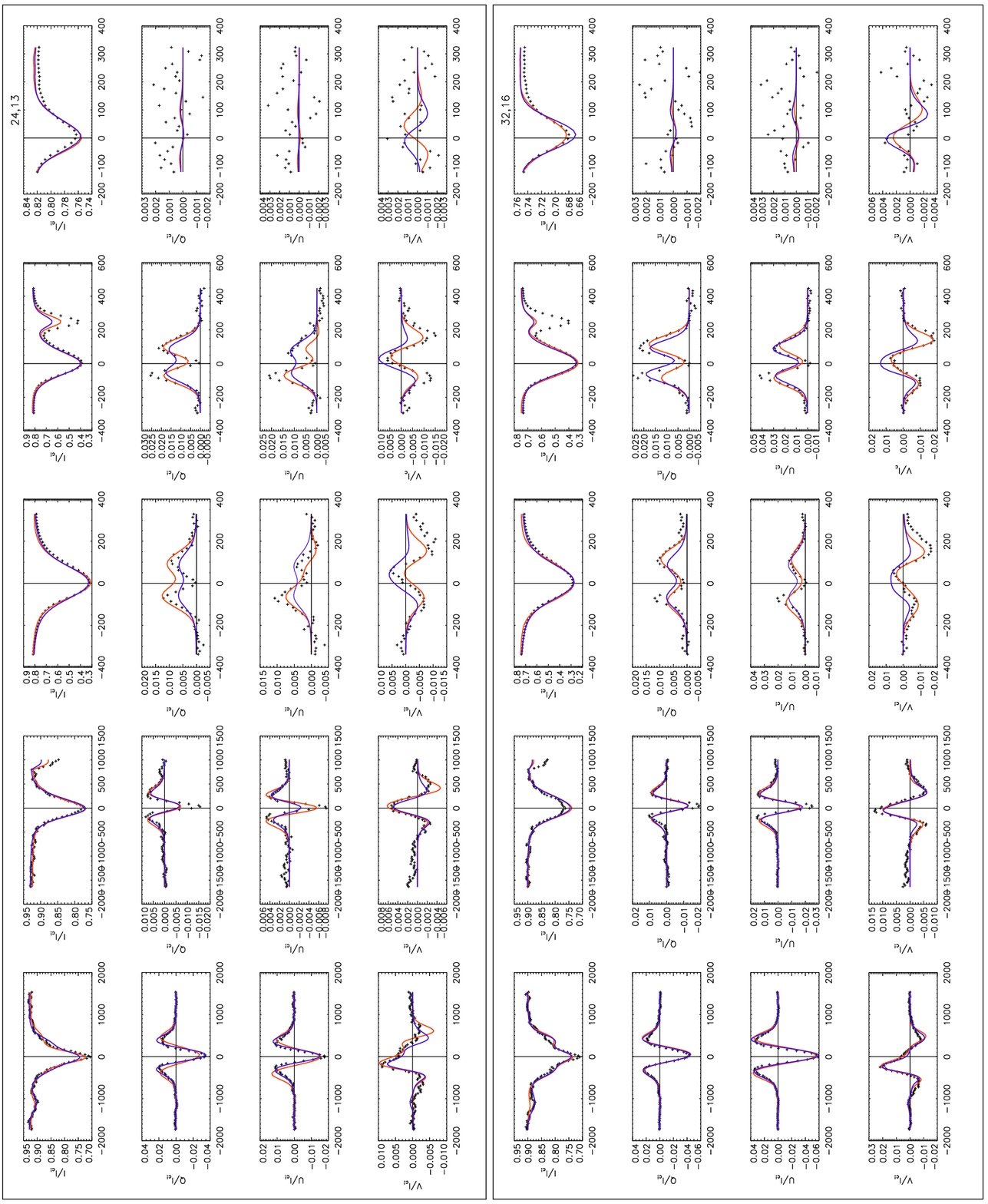}}}
\caption{Profile examples showing the observed spectra ({\em black crosses}),
  the best-fit of the 2C inversion ({\em blue line}), and the best-fit of the
  Gaussian inversion ({\em red line}). In each panel, Stokes $IQUV$ are
    shown from {\em left to right}, and the lines 1564.8\,nm, 1565.2\,nm,
    630.15\,nm, 630.25\,nm, and 630.37\,nm from {\em bottom to top}. The
    locations of the profiles inside the FOV are marked by crosses in Fig.~\ref{fig1}.  \label{prof1}}
\end{figure*}

\begin{figure*}
\centerline{\bf\large 3 \hspace*{8.5cm} 4}$ $\\
\centerline{\resizebox{17.6cm}{!}{\includegraphics*[0,160][523,796]{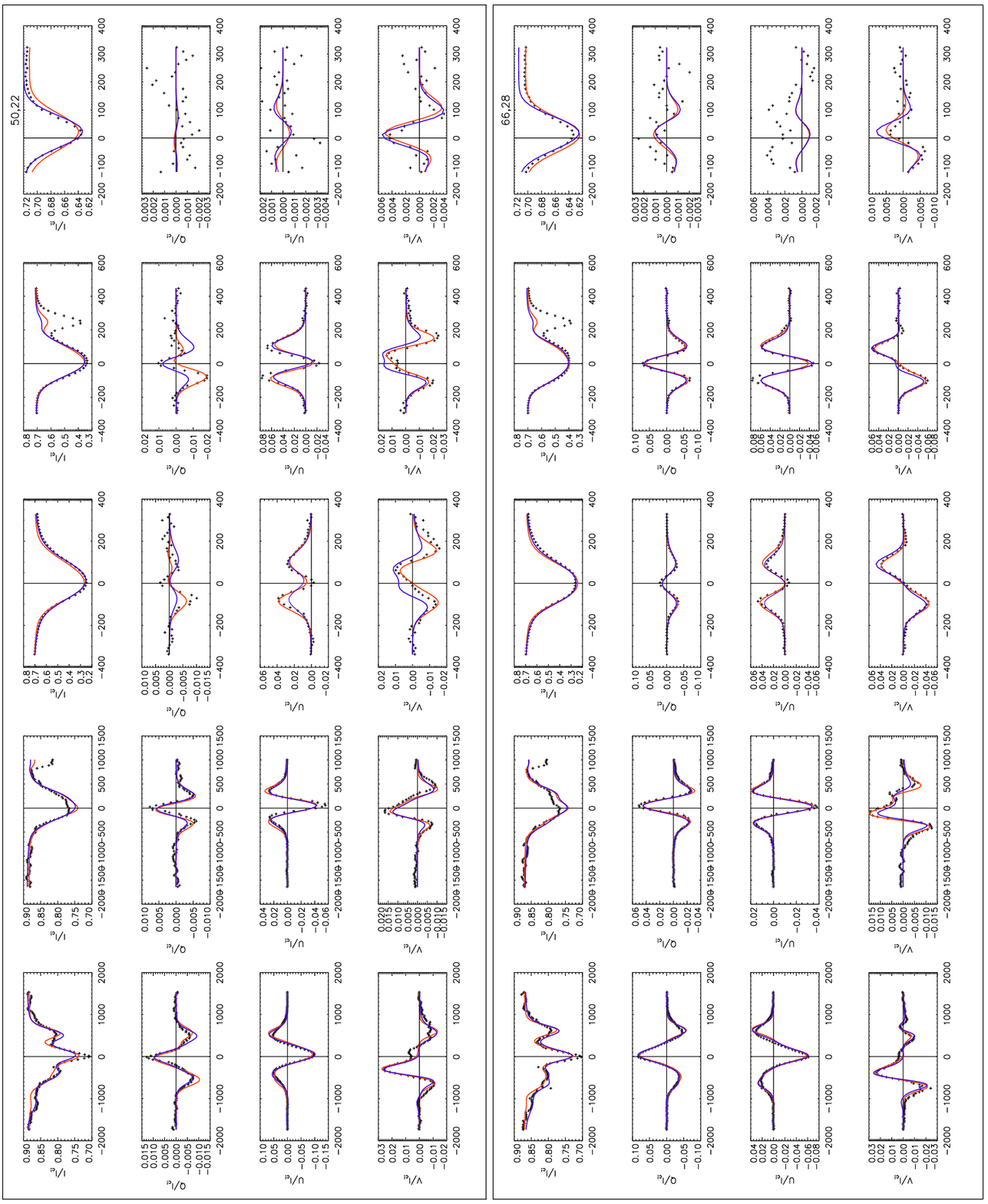}}}
\caption{Same as Fig.~\ref{prof1} for two other pixels. \label{prof2}}
\end{figure*}

\begin{figure*}
\centerline{\bf\large 5 \hspace*{8.5cm} 6}$ $\\
\centerline{\resizebox{17.6cm}{!}{\includegraphics*[0,160][523,796]{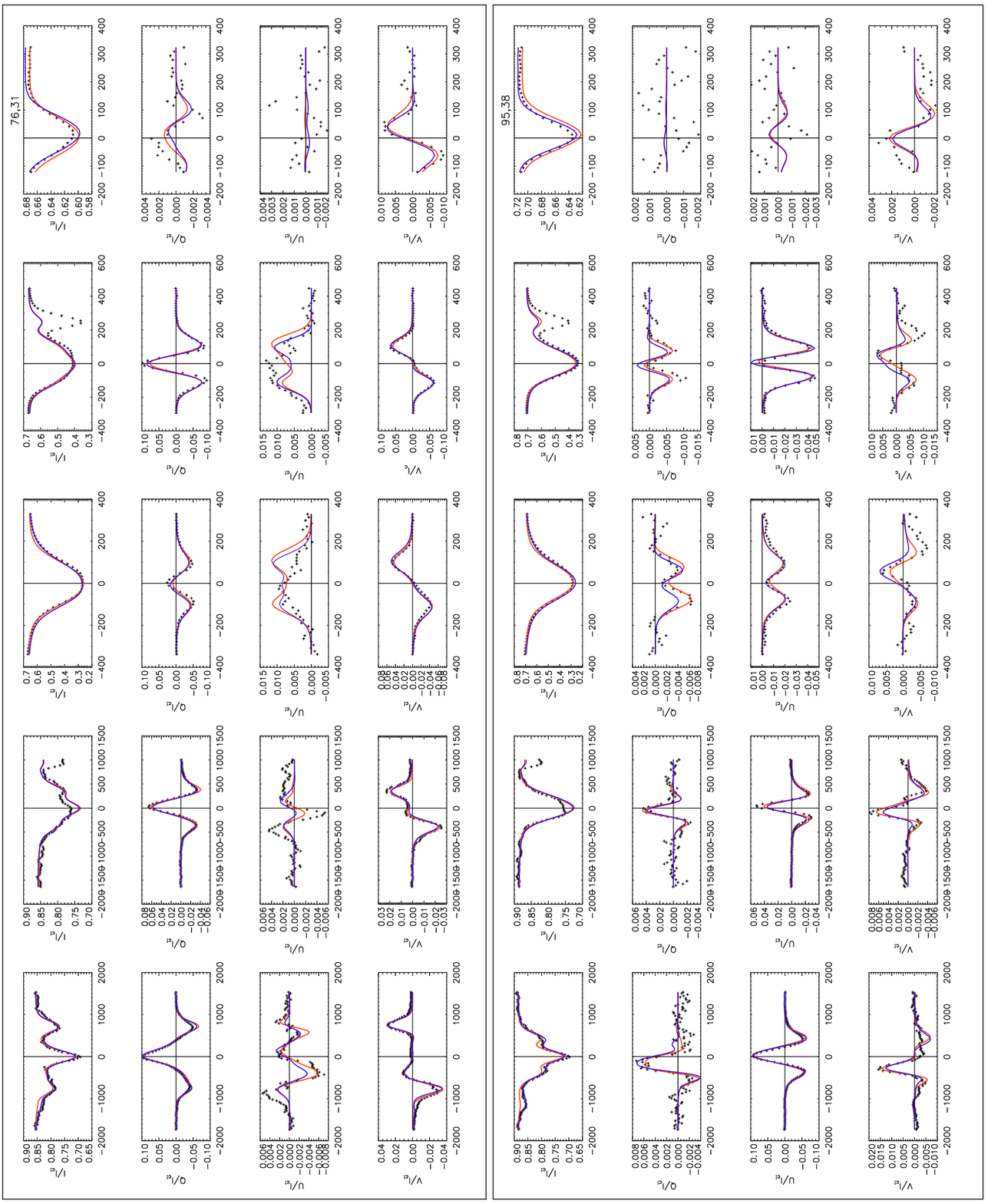}}}
\caption{Same as Fig.~\ref{prof1} for two other pixels. \label{prof3}}
\end{figure*}

Figures \ref{prof1} to \ref{prof3} show the spectra of six different locations
inside the penumbra, marked by crosses in Fig.~\ref{fig1}. The best-fit
profiles of the 2C inversion ({\em blue lines}) can be seen to deviate
strongest from the observations in the case of Stokes $V$ of the VIS lines
like for instance in the {\em left panel} of Fig.~\ref{prof2} (profile no.~3), while still well reproducing the near-IR $V$ spectra at the same time. 

\begin{figure*}
\centerline{\resizebox{17.6cm}{!}{\includegraphics*[46,211][551,773]{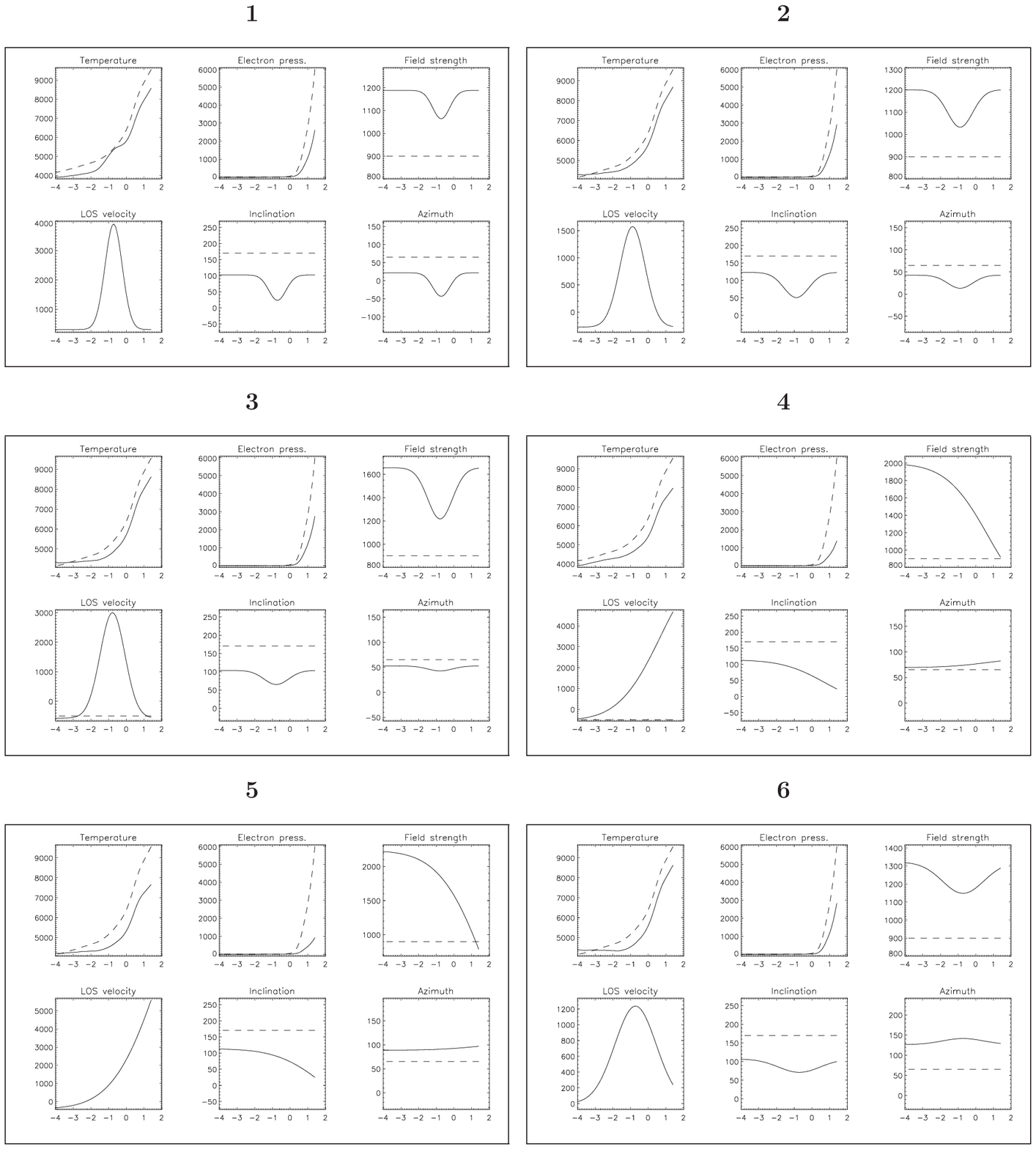}}}

\caption{Atmospheric stratifications of the best-fit result of the Gaussian
  inversion for the profiles shown in Figs.~\ref{prof1} to \ref{prof3}. The layout of each panel is identical to that of Fig.~\ref{gausmodel}. \label{fig_strat}} 
\end{figure*}

The best-fit atmosphere models in Fig.~\ref{fig_strat} show that in some
cases (profiles no.~4 and 5) the Gaussian perturbation actually is converted to a shape quite different to a Gaussian, i.e., an extremely broad Gaussian where the lower boundary of the perturbation is located far below log$\tau$ = 0. In these cases, the 2C inversion setup with constant field parameters and the Gaussian inversion are actually as good as identical approaches to analyze the spectra, and hence yield very similar best-fit spectra.

\end{appendix}
\end{document}